\RequirePackage{lineno}

\documentclass[aps,prl,preprint,superscriptaddress]{revtex4}

\usepackage[dvips]{graphicx,color} 
\usepackage{array,hhline,dcolumn} 
\usepackage{rotating} 

\newcommand{\gev}{\,\mathrm{GeV}}

\newcommand{\dmsq}{\Delta m^{2}}

\newcommand{\numu}{\nu_{\mu}}
\newcommand{\numubar}{\bar{\nu}_{\mu}}

\newcommand{\stwot}{\sin^2 2 \theta}

\newcommand{\evsq}{{eV}^{2}}
\newcommand{\qsq}{{Q}^{2}}

\newcommand{\gtwid}{\mathrel{\raise.3ex\hbox{$>$\kern-.75em\lower1ex\hbox{$\sim$}}}}
\newcommand{\ltwid}{\mathrel{\raise.3ex\hbox{$<$\kern-.75em\lower1ex\hbox{$\sim$}}}}

\begin{document}

%

\title{A search for muon neutrino and antineutrino disappearance in MiniBooNE}
\author{A.~A. Aguilar-Arevalo$^{14}$, C.~E.~Anderson$^{19}$,
	A.~O.~Bazarko$^{16}$, S.~J.~Brice$^{7}$, B.~C.~Brown$^{7}$,
        L.~Bugel$^{5}$, J.~Cao$^{15}$, L.~Coney$^{5}$,
        J.~M.~Conrad$^{5,13}$, D.~C.~Cox$^{10}$, A.~Curioni$^{19}$,
        Z.~Djurcic$^{5}$, D.~A.~Finley$^{7}$, B.~T.~Fleming$^{19}$,
        R.~Ford$^{7}$, F.~G.~Garcia$^{7}$,
        G.~T.~Garvey$^{11}$, C.~Green$^{7,11}$, J.~A.~Green$^{10,11}$,
        T.~L.~Hart$^{4}$, E.~Hawker$^{3,11}$,
        R.~Imlay$^{12}$, R.~A. ~Johnson$^{3}$, G.~Karagiorgi$^{13}$,
        P.~Kasper$^{7}$, T.~Katori$^{10}$, T.~Kobilarcik$^{7}$,
        I.~Kourbanis$^{7}$, S.~Koutsoliotas$^{2}$, E.~M.~Laird$^{16}$,
        S.~K.~Linden$^{19}$, J.~M.~Link$^{18}$, Y.~Liu$^{15}$,
        Y.~Liu$^{1}$, W.~C.~Louis$^{11}$,
        K.~B.~M.~Mahn$^{5}$, W.~Marsh$^{7}$, G.~McGregor$^{11}$,
        W.~Metcalf$^{12}$, P.~D.~Meyers$^{16}$,
        F.~Mills$^{7}$, G.~B.~Mills$^{11}$,
        J.~Monroe$^{5}$, C.~D.~Moore$^{7}$, R.~H.~Nelson$^{4}$,
	V.~T.~Nguyen$^{5,13}$, P.~Nienaber$^{17}$, J.~A.~Nowak$^{12}$,
	B.~Osmanov$^{8}$, S.~Ouedraogo$^{12}$, R.~B.~Patterson$^{16}$,
        D.~Perevalov$^{1}$, C.~C.~Polly$^{9,10}$, E.~Prebys$^{7}$,
        J.~L.~Raaf$^{3}$, H.~Ray$^{8,11}$, B.~P.~Roe$^{15}$,
	A.~D.~Russell$^{7}$, V.~Sandberg$^{11}$, R.~Schirato$^{11}$,
        D.~Schmitz$^{5}$, M.~H.~Shaevitz$^{5}$, F.~C.~Shoemaker$^{16}$,
        D.~Smith$^{6}$, M.~Soderberg$^{19}$,
        M.~Sorel$^{5}$\footnote{Present address: IFIC, Universidad de Valencia and CSIC, Valencia 46071, Spain},
        P.~Spentzouris$^{7}$,J.~Spitz$^{19}$,I.~Stancu$^{1}$,
        R.~J.~Stefanski$^{7}$, M.~Sung$^{12}$, H.~A.~Tanaka$^{16}$,
        R.~Tayloe$^{10}$, M.~Tzanov$^{4}$,
        R.~Van~de~Water$^{11}$, 
	M.~O.~Wascko$^{12}$\footnote{Present address: Imperial College; London SW7 2AZ, United Kingdom},
	 D.~H.~White$^{11}$,
        M.~J.~Wilking$^{4}$, H.~J.~Yang$^{15}$,
        G.~P.~Zeller$^{5,11}$, E.~D.~Zimmerman$^{4}$ \\
\smallskip
(The MiniBooNE Collaboration)
\smallskip
}
\smallskip
\smallskip
\affiliation{
$^1$University of Alabama; Tuscaloosa, AL 35487 \\
$^2$Bucknell University; Lewisburg, PA 17837 \\
$^3$University of Cincinnati; Cincinnati, OH 45221\\
$^4$University of Colorado; Boulder, CO 80309 \\
$^5$Columbia University; New York, NY 10027 \\
$^6$Embry Riddle Aeronautical University; Prescott, AZ 86301 \\
$^7$Fermi National Accelerator Laboratory; Batavia, IL 60510 \\
$^8$University of Florida; Gainesville, FL 32611 \\
$^9$University of Illinois; Urbana, IL 61801 \\
$^{10}$Indiana University; Bloomington, IN 47405 \\
$^{11}$Los Alamos National Laboratory; Los Alamos, NM 87545 \\
$^{12}$Louisiana State University; Baton Rouge, LA 70803 \\
$^{13}$Massachusetts Institute of Technology; Cambridge, MA 02139 \\
$^{14}$Universidad National Autonoma de Mexico, Mexico, D.F., Mexico \\
$^{15}$University of Michigan; Ann Arbor, MI 48109 \\
$^{16}$Princeton University; Princeton, NJ 08544 \\
$^{17}$Saint Mary's University of Minnesota; Winona, MN 55987 \\
$^{18}$Virginia Polytechnic Institute \& State University; Blacksburg, VA
24061
\\
$^{19}$Yale University; New Haven, CT 06520\\
}

\date{\today}

\begin{abstract}
 The MiniBooNE Collaboration reports a search for $\numu$ and $\numubar$ disappearance in the $\dmsq$ region of a few $\evsq$. These measurements are important for constraining models with extra types of neutrinos, extra dimensions and CPT violation. Fits to the shape of the $\numu$ and $\numubar$ energy spectra reveal no evidence for disappearance at  90\% confidence level (CL) in either mode. This is the first test of $\numubar$ disappearance between $\dmsq=0.1-10~\evsq$.
\end{abstract}

\pacs{14.60.Lm, 14.60.Pq, 14.60.St}

\keywords{Suggested keywords}
\maketitle

Neutrino oscillations have been observed and confirmed at mass splittings ($\dmsq$) of $\sim 10^{-5}~\evsq$ and $\sim 10^{-3}~\evsq$, called the ``solar" and ``atmospheric" oscillations respectively. The observed mixing is consistent with three generations of neutrinos and a unitary mixing matrix. Complicating this picture, the LSND experiment observed an excess of  $\overline{\nu}_e$ in a $\overline{\nu}_{\mu}$ beam \cite{lsnd}, indicating a possible third $\dmsq$ around $1 ~\evsq$ thus requiring more than three neutrino generations or other exotic physics. Recently, the MiniBooNE experiment \cite{mb_osc} excluded two neutrino appearance-only oscillations (98\% CL) as an explanation of the LSND excess if oscillations of neutrinos and antineutrinos are the same.

Exotic physics models \cite{weiler,karagiorgi,maltoni,nelson}, including sterile neutrinos, extra dimensions, and CPT violation have been proposed to explain the LSND observation. Some of these models can also accommodate the MiniBooNE $\nu_e$ appearance oscillation results. These models are testable with measurements of $\numu$ and  $\overline{\nu}_{\mu}$ disappearance which constrain any non-standard oscillations of  $ \stackrel{ \scriptscriptstyle ( - ) }{\numu} \rightarrow  \stackrel{ \scriptscriptstyle ( - ) }{\nu_x} $. As described in this Letter, the MiniBooNE experiment has performed searches for  $\nu_{\mu}$ and  $\overline{\nu}_{\mu}$ disappearance which probe a region of interest, $\dmsq=0.5-40 ~\evsq$, not covered by two previous disappearance experiments, CCFR  ($\nu_{\mu}$ and  $\overline{\nu}_{\mu}$)\cite{Stockdale:1984cg} and CDHS ($\nu_{\mu}$ only)\cite{Dydak:1983zq}. Unless otherwise stated, all statements about neutrinos hold true also for antineutrinos. 

For the MiniBooNE experimental setup, the detector is located at a fixed distance from the neutrino source. In this case, $\numu$ disappearance due to oscillations has a distinct signature as a function of neutrino energy, because neutrinos with different energies oscillate with different probabilities for the same distance traveled. Disappearance would be observable either via a deficit of events (normalization) or, alternatively, via a distortion of the neutrino energy spectrum (shape), or both (normalization + shape). The absolute normalization uncertainties in a single detector experiment such as MiniBooNE are large, hence a shape-only disappearance fit is performed.
The $\stackrel{ \scriptscriptstyle ( - ) }{\numu}$ flux to the MiniBooNE detector is provided by the Fermilab Booster Neutrino Beam (BNB) which is produced by 8 GeV protons incident on a 1 cm diameter, 71 cm long (1.7 interaction length) beryllium target  surrounded by a  magnetic horn pulsed at 174 kA. The horn uses positive current to focus $\pi^+$ and $K^+$  mesons for the neutrino mode sample and negative current to focus $\pi^-$ and $K^-$ for the antineutrino mode sample. The mesons that pass through a 60 cm diameter collimator 259 cm downstream of the target decay in a 50 m long tunnel to produce the $\stackrel{ \scriptscriptstyle ( - ) }{\numu}$ beam. The BNB flux \cite{mb_flux} is determined using a GEANT4\cite{geant4} based beam simulation which has been further modified to include updated p-Be particle production data.

The distance from the proton interaction target to the MiniBooNE detector \cite{mb_detector} is 541 m. The MiniBooNE detector is a 12 m diameter spherical tank filled with 800 tons of mineral oil. The detector is separated into an inner region filled with 1280 inward facing 8 inch photomultiplier tubes (PMTs) and an optically isolated outer region used to reject cosmic-ray induced events. Charged particles produced in neutrino interactions emit primarily Cherenkov light, though a small amount of scintillation light is also produced. Light and particle production and propagation in the MiniBooNE detector is modeled using a GEANT3 \cite{geant3} based simulation, which was tuned using MiniBooNE and external data. 

Neutrino interactions are simulated with the v3 NUANCE event generator \cite{nuance}.Prior to selection, approximately 42\% of all events in MiniBooNE are charged current quasi-elastic (CCQE) scattering and 22\% are charged current single charged pion production (CC$1\pi^{+/-}$) in both neutrino and antineutrino mode.

The search for oscillations is conducted with a sample of CCQE events because of the high statistics and purity. The reconstructed neutrino energy ($E_{\nu}^{QE}$) is calculated assuming the target nucleon is at rest:

\begin{equation}
E_{\nu}^{QE} = \frac{2(M_n-E_B)E_{\mu}-(E_{B}^2-2M_nE_B + \Delta{M} + {M^2}_{\mu})}{2[(M_n-E_B)-E_{\mu} + p_{\mu}cos\theta_{\mu}]}
\end{equation}

\noindent
where $\Delta{M}=M_{n}^2-M_{p}^2$, M indicates the muon, proton or neutron mass with appropriate subscripts, $E_{B}$ is the nucleon binding energy, $E_{\mu}$($p_{\mu}$) is the reconstructed muon energy (momentum) and $\theta_{\mu}$ is the reconstructed muon scattering angle with respect to the neutrino beam direction. A small correction is applied in both data and simulation to account for the biasing effects of Fermi-smearing. At 300 MeV, the muon energy resolution is 7\% and the angular resolution is 5 degrees. The average $E_{\nu}^{QE}$ resolution is 11\% for CCQE events \cite{mb_qe}.

A CCQE event sample is selected by identifying a single muon in the detector and its associated decay electron using the same criteria as in the previous measurement of CCQE model parameters on carbon \cite{mb_qe}. Timing information from the PMTs allows the light produced by the initial neutrino interaction (first ``subevent'') to be separated from the light produced by the decay electron (second ``subevent''). The timing and charge response of the PMTs is then used to reconstruct the position, kinetic energy and direction vector of the primary particle within each subevent. Exactly two subevents are required in the analysis (the muon and its decay electron). To reject cosmic ray interactions, both subevents are required to have fewer than 6 veto-PMT hits. The first subevent must be in coincidence with a beam pulse, have a reconstructed track center less than 500cm, and greater than 200 inner tank PMT hits to eliminate electrons from stopped cosmic ray muon decays. The second subevent must have less than 200 inner PMT hits to be below the decay electron energy endpoint. Finally, the distance between the electron vertex and the muon track endpoint must be less than 100 cm, ensuring that the electron is associated with the muon track. This selection also applies to the antineutrino mode sample as the final state nucleon is not reconstructed and the detector does not distinguish muon charge.

The selection yields 190,454 data events with $0 < E_{\nu}^{QE} < 1.9$ GeV for $5.58 \times 10^{20}$ protons on target (POT) in the neutrino mode sample; 27,053 data events for $3.39 \times 10^{20}$ POT in the antineutrino mode sample. According to the simulation, the neutrino mode sample is 74\% pure CCQE, and the antineutrino mode sample is 70\% pure CCQE. The primary background ($\sim$75\%) for both the $\numu$ and $\numubar$ samples is CC$1\pi$ events where the outgoing pion is unobserved (e.g. due to absorption in the nucleus). Though the neutrino mode sample has $<1$\% $\numubar$ content, in antineutrino mode, the beam contains a substantial contribution of $\numu$ due to the higher $\pi^{+}$ production at the target and the higher $\numu$ cross section. The antineutrino mode is predicted to have 25\% $\numu$ content.

The CCQE cross section depends on the axial vector form factor, which is commonly assumed to have a dipole form as a function of four-momentum transfer ($\qsq$) with one adjustable parameter, $M_A$, the axial mass. Global fits to the world's neutrino scattering data on deuterium yield $M_A=1.015 \gev$~\cite{Bodek:2007vi}, however recent results from K2K \cite{Espinal:2007zz,Gran:2006jn} and MiniBooNE \cite{mb_qe} suggest a higher effective value of $M_A\sim1.2$ for nuclear targets. In addition, MiniBooNE has also adjusted the level of Pauli blocking in the prediction, using a parameter $\kappa=1.019$, to better reproduce the experimental data at low $\qsq$~\cite{mb_qe}. The effect of $M_A$ and $\kappa$ on the $\qsq$ shape is pronounced while oscillations provide relatively little $\qsq$ distortion. Therefore, the effect of the cross section tuning does not mask any underlying disappearance in the neutrino or antineutrino mode samples.

For the disappearance search, systematic uncertainties are included for the underlying neutrino flux prediction, neutrino interaction cross section, and detector response. The method used to estimate the uncertainties due to the underlying neutrino flux prediction and detector model is identical to that used in previous MiniBooNE results \cite{mb_osc,lowe}.  The dominant uncertainty on the CCQE cross section is from uncertainties on $M_A$ and $\kappa$, which span the difference between the deuterium and nuclear target results ($M_A=1.015\pm0.20\gev$, $\kappa=1.000 \pm 0.019$). In addition, the uncertainty on the shape of the CC$1\pi$ background is estimated using the MiniBooNE CC$1\pi^+$ data sample.  With $M_A=1.015 \gev$, the ratio of detected events to predicted events in MiniBooNE for neutrinos is $1.31\pm0.26$, $1.18\pm0.18$ for antineutrinos, which shows agreement within the uncertainties. The difference between this value and previously published values is due to the different values of $M_A$ and $\kappa$~\cite{mb_qe}. Systematic uncertainties produce correlated errors between  $E_{\nu}^{QE}$ bins that are included by developing a covariance matrix in the same manner as in previous MiniBooNE oscillation analyses \cite{mb_osc,lowe}. This covariance matrix includes separate normalization and shape-only error contributions. For the shape-only disappearance search, the prediction is normalized to data, and just the shape-only covariance matrix is used.

\begin{figure}[htbp]
\centerline{\includegraphics[height=3.5in,clip = true]{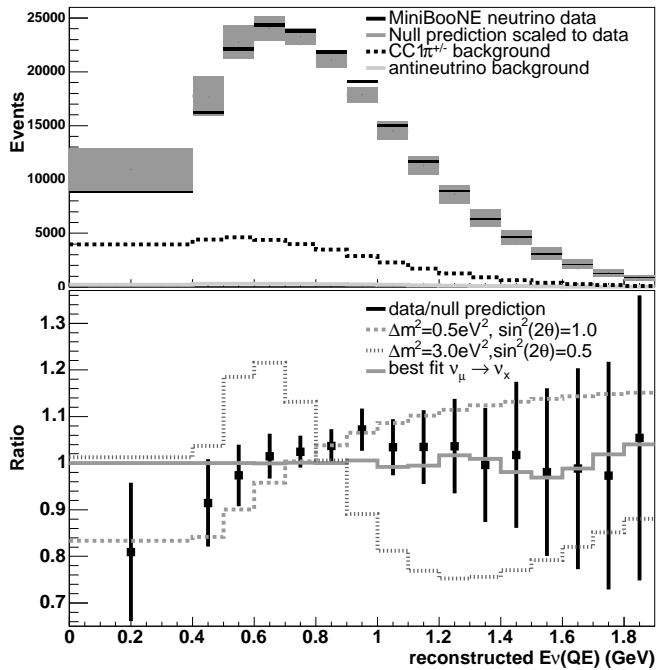}}
\vspace{-.15in}
\caption{The top plot shows the $E_{\nu}^{QE}$ distribution for neutrino data (black) with statistical error rectangles (thickness of line indicates size of statistical error), and prediction assuming no oscillation (grey). Attached to the prediction are the diagonal elements of the shape error matrix. The predicted CC$1\pi$ background (dash) and background antineutrino (solid) events are also shown. The bottom plot shows the ratio of data to no oscillation (black), and the ratio of no oscillation to: $\dmsq=0.5 ~\evsq$, $\stwot=1.0$ disappearance (dashed line), $\dmsq=3.0 ~\evsq$, $\stwot=0.5$ disappearance (dotted line) and for the minimum $\chi^2=12.72$ (13 DF) at $\dmsq=17.5 ~\evsq$, $\stwot=0.16$ (solid line). }
\label{datamc_numu}
\end{figure}

\begin{figure}[htbp]
\centerline{\includegraphics[height=3.5in,clip = true]{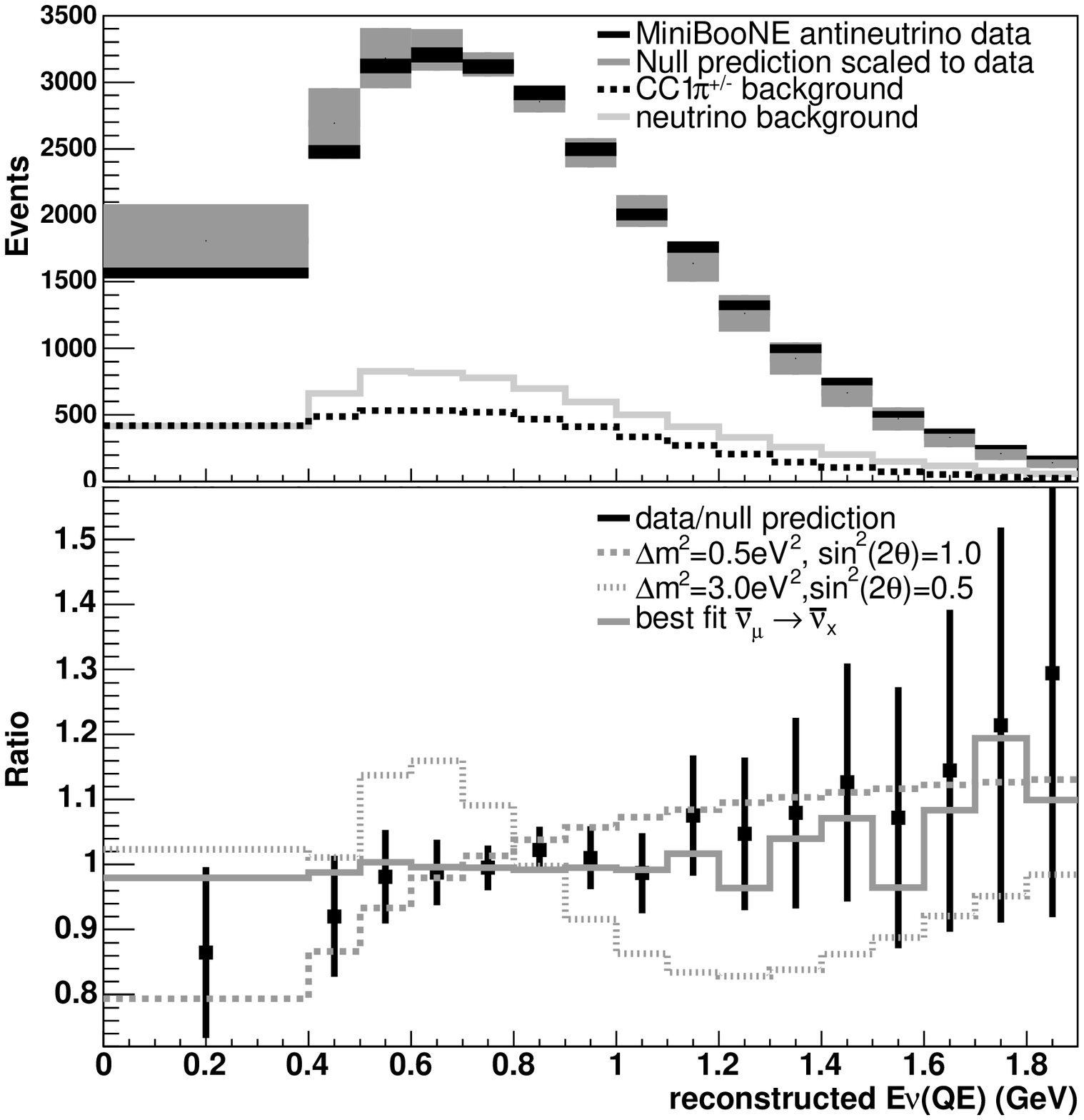}}
\vspace{-.15in}
\caption{Same convention as Fig. \ref{datamc_numu} for the antineutrino mode sample. Minimum $\chi^2=5.43$ (11 DF) is at $\dmsq=31.3 ~\evsq$, $\stwot=0.96$. }
\label{datamc_numubar}
\end{figure}


\begin{figure}[htbp]
\centerline{\includegraphics[scale=0.5,clip = true]{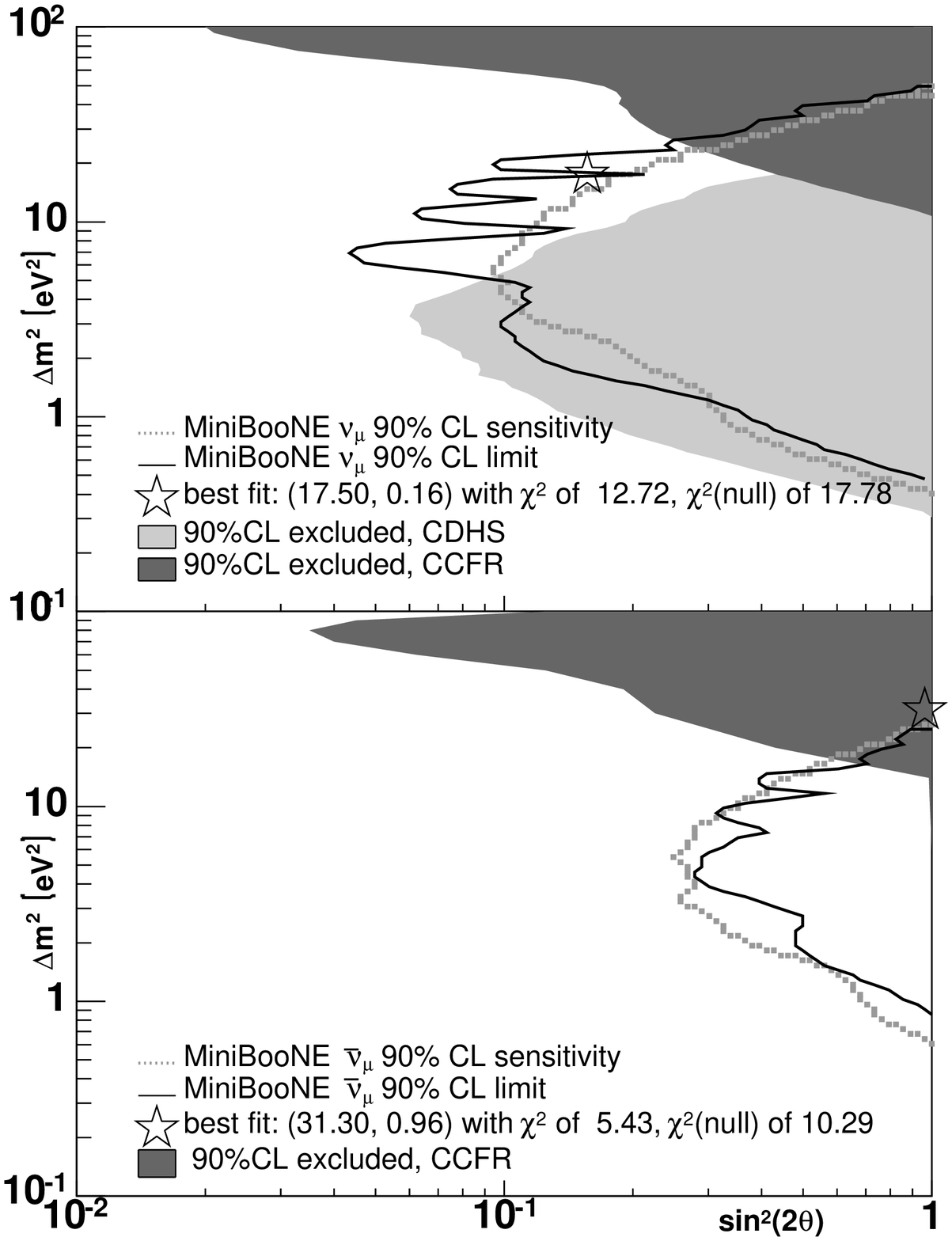}}
\vspace{-.15in}
\caption{The top plot shows the sensitivity (dashed line) and limit (solid line) for 90\% CL for neutrino disappearance in MiniBooNE. Previous limits by CCFR (dark grey) and CDHS (light grey) are also shown. The bottom plot uses the same convention for  antineutrino disappearance. }
\label{limit}
\end{figure}

The disappearance search uses the Pearson's $\chi^2$ test to determine allowed regions in the $\dmsq-\stwot$ plane. The $\chi^2$ is calculated from a comparison of the data, $d_i$, in $E_{\nu}^{QE}$ bin $i$ to a prediction $p_i(\dmsq,\stwot)$ for 16 bins. The prediction assumes two-flavor $\numu \rightarrow \nu_x$ disappearance characterized by one large mass splitting ($\dmsq \equiv \Delta m^2_{hk}$) between the light neutrino mass states $k$, which participate in standard three neutrino oscillation, and $h$, the heavier neutrino state, and one oscillation amplitude $\stwot=4{|U_{\mu,h}|}^2(1-{|U_{\mu,h}|}^2)$, where ${|U_{\mu,h}|}^2$ is the muon flavor content of the heavy state $h$.
\vspace{-.08in}
\begin{equation}
\chi^2 = \sum_{i,j}^{16 bins}(d_i-Np_i) {M_{ij}}^{-1}(d_j-Np_j) 
\end{equation}\vspace{-.12in}

\noindent
where $M_{ij}$ is the shape-only error matrix, and $N$ is a factor which normalizes the prediction to the total number of observed events in data. All neutrino events in the prediction, including the  CC$1\pi^+$ background events, are allowed to oscillate in the fit based on the incident neutrino energy and distance traveled. The 90\%CL limit corresponds to $\chi^2>23.5$ for 16 degrees of freedom (DF). The sensitivity is a fit to an unoscillated prediction including all statistical and systematic uncertainties. 

The top plot of Fig. \ref{datamc_numu} shows $E_{\nu}^{QE}$ after selection cuts for the neutrino data and the prediction assuming no oscillation (null hypothesis) with diagonal elements of the error matrix.  The dominant systematics arise from the neutrino flux (production of $\pi^{+/-}$ from p-Be interactions) and CCQE cross section; uncertainties at low energy are larger because of the substantial CC$ 1\pi^+$ background and uncertainties in the CCQE cross section in this region.  Though the diagonal elements of the error matrix are substantial, the correlations between energy bins are large. The $\chi^2$ between the data and the null hypothesis is 17.78 (16 DF, 34\% probability) for the neutrino mode sample which is consistent with no oscillation at 90\%CL. The top plot of Fig. \ref{limit} shows the 90\% CL sensitivity and limit curves for the neutrino mode sample. The minimum $\chi^2=12.72$ (13 DF, 47\% probability) at $\dmsq=17.5 ~\evsq$, $\stwot=0.16$, where the number of degrees of freedom is estimated from frequentist studies.

The bottom plot in Fig. \ref{datamc_numu} shows the ratio of data to the null hypothesis and three oscillation scenarios. The shape distortion for $\dmsq=0.5~\evsq$ is very different from  $\dmsq=3.0~\evsq$.  The $\chi^2$ therefore changes rapidly as a function of $\dmsq$, resulting in rapid changes in the 90\%CL sensitivity and limit curves (Fig. \ref{limit}) for small differences in $\dmsq$\hspace{-0.07in}. Similar features are also seen in previous disappearance analyses \cite{Stockdale:1984cg,Dydak:1983zq}.

The $\numubar$ disappearance analysis proceeds in the same manner as the $\numu$ analysis, except that only the $\numubar$ events are allowed to oscillate in the fit and the $\numu$ events are kept fixed. This determines the limit on a model where the $\numubar$ can oscillate but the $\numu$ cannot. A model where both $\numu$ and $\numubar$ oscillate with equal oscillation probability versus energy would produce a limit very similar to the neutrino mode limit.

During antineutrino data taking, two absorber plates inadvertently fell vertically into the decay volume at 25m and were later removed, creating three distinct data taking periods with zero, one, or two absorbers in the beamline. The event rate was predicted to be 13\% (20\%) lower for one (two) plate(s) in the beam. Approximately 15\% of the antineutrino data taken had one absorber plate inserted, and 15\% had two absorber plates inserted. Because the changes to the beamline are understood, a separate simulation was run with the appropriate number of absorber plates in the beamline. Figure \ref{datamc_numubar} shows the  $E_{\nu}^{QE}$ distribution for the antineutrino mode sample. The $\chi^2$ of the null hypothesis is 13.7, 8.2, 15.2, 10.29 (16 DF) for the zero, one, and two absorber plate and total data respectively. The antineutrino mode data is also consistent with no oscillation at 90\%CL, so the bottom plot of  Fig. \ref{limit} shows the 90\% CL sensitivity and limit curves for the antineutrino disappearance fit to all antineutrino data; the limit curves for the individual absorber data periods were found to be consistent with the total. In addition to the two-neutrino oscillation fits described above, some studies were performed some of the MiniBooNE energy spectra within the context of 3+2 oscillation models. The best fits for 3+2 sterile neutrino models in Ref. \cite{karagiorgi} are consistent with the MiniBooNE $\numu$ and  $\numubar$ data, but the $\numu$ data rules out the best fit point from the global fit to MiniBooNE $\nu_e$ data in Ref. \cite{maltoni} at 90\% CL with $\chi^2=24.7$(16 DF).

In summary, MiniBooNE observes no evidence for $\numu$ or $\numubar$ disappearance at 90\%CL in the $\dmsq$ region of a few $\evsq$. The test of $\numubar$ disappearance probes a region of $\dmsq=0.1-10~\evsq$ unexplored by previous experiments.

\begin{acknowledgments}
We acknowledge the support of Fermilab, the Department of Energy,
and the National Science Foundation. 
We thank Los Alamos National Laboratory for LDRD funding.
 We also acknowledge the use of Condor software in the analysis of the data.
\end{acknowledgments}


\end{document}